\documentclass[12pt]{article}

\newsavebox{\ns}
\newsavebox{\dbrane}
\newsavebox{\dbshort}

\usepackage{epsfig}
\usepackage{amssymb}

\def\be{\begin{eqnarray}}
\def\ee{\end{eqnarray}}

\newcommand{\nn}{\nonumber}
\newcommand\para{\paragraph{}}

\newcommand{\eqn}[1]{(\ref{#1})}

\newcommand\vkn{${\cal V}_{k,N_c}$}

\def\Dslash{\,\,{\raise.15ex\hbox{/}\mkern-12mu D}}
\def\Dbarslash{\,\,{\raise.15ex\hbox{/}\mkern-12mu {\bar D}}}
\def\delslash{\,\,{\raise.15ex\hbox{/}\mkern-9mu \partial}}
\def\delbarslash{\,\,{\raise.15ex\hbox{/}\mkern-9mu {\bar\partial}}}
\def\pslash{\,\,{\raise.15ex\hbox{/}\mkern-9mu p}}
\def\calDslash{\,\,{\raise.15ex\hbox{/}\mkern-12mu {\cal D}}}

\begin{document}
\pagestyle{plain}
\setcounter{page}{1}
\newcounter{bean}
\baselineskip16pt

\begin{titlepage}

\begin{center}
\today
\hfill hep-th/0403158\\
\hfill MIT-CTP-3483 \\

\vskip 1.5 cm
{\large \bf Vortex Strings and Four-Dimensional Gauge Dynamics}
\vskip 1 cm 
{Amihay Hanany and David Tong}\\
\vskip 1cm
{\sl Center for Theoretical Physics, 
Massachusetts Institute of Technology, \\ Cambridge, MA 02139, U.S.A.\\} 
{\tt hanany, dtong@mit.edu}

\end{center}

\vskip 0.5 cm
\begin{abstract}
We study the low-energy quantum dynamics of vortex strings in the Higgs phase of 
${\cal N}=2$ supersymmetric QCD. The exact BPS spectrum of the stretched string is 
shown to coincide with the BPS spectrum of the four-dimensional parent gauge theory. 
Perturbative string excitations correspond to bound W-bosons and quarks 
while the monopoles 
appear as kinks on the vortex string. This provides a physical 
explanation for an observation by N. Dorey relating 
the quantum spectra of theories in two and four dimensions. 

\end{abstract}

\end{titlepage}

\section{Introduction and Conclusion}

Two-dimensional sigma-models have long acted as a playground in which 
to test aspects of four-dimensional gauge dynamics. The two systems 
share many qualitative features including asymptotic freedom, a 
dynamically generated mass gap, anomalies,  and instantons.

\para
Some years ago, N. Dorey proved a more quantitative correspondence between 
supersymmetric theories in two and four dimensions \cite{nick}. He showed 
that the BPS spectrum of the mass deformed two-dimensional 
${\cal N}=(2,2)$ ${\bf CP}^{N-1}$ sigma-model coincides with the BPS 
spectrum of  four-dimensional ${\cal N}=2$ $SU(N)$ supersymmetric QCD. The 
correspondence is exact, holding at the quantum level in both weak and strong 
coupling regimes. Generalisations to other two-dimensional sigma-models 
were later found \cite{dht}. However, despite some insight from brane 
constructions \cite{hh,dht},  
the underlying reason for the agreement remained mysterious. The purpose of 
this paper is to provide a field theoretic explanation for the correspondence.

\para 
The key to our story lies in the recent progress in understanding 
the dynamics of various soliton configurations in the Higgs phase of 
${\cal N}=2$ SQCD. Of particular relevance for our story are the non-abelian vortices 
\cite{vib}, which are string-like objects in four dimensions, and monopoles 
\cite{monflux} which, due to the Meissner effect, are confined in the Higgs phase and 
come attached to two semi-infinite vortex strings\footnote{Analogous configurations in a 
closely related theory were also discussed in \cite{yung} and \cite{auzzi} respectively. 
Other work on confined monopoles can be found in \cite{no}}. 
We shall show that the two-dimensional 
theory considered by Dorey in \cite{nick} is precisely the theory describing 
the vortex string. As we explain below, the BPS excitations of the string 
have an interpretation as four dimensional states: the perturbative string 
excitations correspond to W boson - string bound states, 
while the solitonic kinks of the string correspond to the confined monopoles in four 
dimensions \cite{monflux}.

\para
The results presented here fit into the growing body of work devoted to 
understanding the dynamics of solitons in the Higgs phase of ${\cal N}=2$ 
theories. In recent years we have found that these theories admit a 
remarkably rich structure of classical BPS solitons. As well as the strings 
and confined monopoles mentioned above, 
there is an intricate system of domain walls \cite{at,more}, domain wall junctions 
\cite{junction} and, perhaps most remarkably, D-branes \cite{dbrane,shif} 
in which the vortex string terminates on a domain wall where its end is electrically 
charged under a gauge field. While field theoretic D-branes have been known to 
exist for some time in strong coupling regimes \cite{dbrane2} the objects described in 
\cite{dbrane,shif} are amenable to semi-classical analysis.

\para
The conclusion of this paper -- that the quantum dynamics of solitons, 
specifically vortex strings, may be used to extract information about the 
strong coupling dynamics of the underlying four dimensional gauge theory --   
is reminiscent of the stringy games played in ten dimensions. For  
example in the the old-new Matrix theory, D-brane solitons 
contain much information about the bulk dynamics. It would be interesting 
to see if this analogy can be pushed further.

\para
The correspondence discovered in \cite{nick} holds in both strong coupling and weak 
coupling regimes of the two theories. In the latter regime, the central charge of 
the theory may be expanded in an infinite series of instanton contributions. 
Since the BPS spectra coincide, this expansion 
agrees term by term, suggesting a quantitative correspondence between two 
dimensional instantons (which are vortices) and four-dimensional Yang-Mills instantons.
Indeed, in \cite{vib}, an ADHM-like construction of the vortex moduli space was 
presented and it was shown that the moduli space of vortices is a particular 
submanifold of the 
moduli space of instantons. It would be interesting to prove explicitly that the 
integrals over the relevant moduli spaces coincide. From the interpretation of the 
correspondence presented here, this agreement suggests another solitonic connection: 
a vortex in a vortex string looks like a Yang-Mills instanton in four dimensions. In Section 
3, we present the Bogomoln'yi equations describing such a solution.

\para
The plan of the paper is as follows. In Section 2, we study 
${\cal N}=2$ supersymmetric $U(N_c)$ gauge theory with $N_f=N_c$ flavours. We 
review the exact central charge on the Coulomb branch which can be determined 
from the Seiberg-Witten solution. We then follow the states as you slide 
onto the Higgs branch, breaking the gauge group completely. 
We shall show that the monopoles remain BPS, but are now confined. At the same 
time, a new BPS object appears: the vortex string. 
In Section 3 we describe the low-energy dynamics of the vortex 
string and show that it coincides with the two-dimensional theory studied 
in \cite{nick}. We review the computation of the BPS spectrum and confirm 
that it does indeed coincide with that of the four-dimensional 
parent theory. In particular, we shall see that elementary excitations of 
the string are associated to W-bosons, while kinks in the string are 
monopoles in four dimensions. 
In Section 4, we repeat this story for $N_f>N_c$ flavours, and the associated 
``semi-local'' vortices, giving a rationale for the generalisation 
discovered in \cite{dht}. This includes the case of the conformal vortex 
string. 

\para
{\bf Note Added:} After finishing this work we were informed that similar 
conclusions have been reached by M. Shifman and A. Yung \cite{misha}. We 
would like to thank M. Shifman for communicating their results to us prior 
to publication.

\section{The Four Dimensional Gauge Theory: $N_f=N_c$}

Our interest in this paper will focus on  ${\cal N}=2$ supersymmetric QCD 
with $U(N_c)$ gauge group with $N_f$ flavours transforming in the fundamental
representation. In this section we restrict to $N_f=N_c$. Nevertheless 
we shall continue to use the subscripts $f$ and $c$ to distinguish between 
flavour and colour groups. Generalisation to $N_f>N_c$ will be given in Section 4. 
We denote the complexified gauge coupling constant
\footnote{A note on conventions: our Yang-Mills term is normalised as 
$(1/4e^2){\rm Tr}(F_{\mu\nu}F^{\mu\nu})$ which differs by a quadratic Casimir 
factor of $2$ from the usual conventions. This leads to an 
unfamiliar factor of 2 in this and other formulae containing $e^2$.} 
as $\tau=2\pi i/e^2 +\theta/2\pi$.

\para
In ${\cal N}=1$ language the theory contains a $U(N_c)$ vector multiplet 
field, an adjoint chiral multiplet $\Phi$ and a further $2N_f$ chiral 
multiplets $Q_i$ and $\tilde{Q}_i$, $i=1,\ldots, N_f$. The $Q_i$ transform 
in the $({\bf N}_c,\bar{\bf N}_f)$ of the $U(N_c)\times SU(N_f)$ gauge and 
flavour group. The $\tilde{Q}_i$ transform in the $(\bar{\bf N}_c,{\bf N}_f)$.
The lowest component of each chiral multiplet is a complex scalar 
field which, as is traditional, we denote by the corresponding lower-case 
letter i.e. $\phi$, $q_i$ and $\tilde{q}_i$.  We 
provide each of the hypermultiplets with a complex mass parameter $m_i$ through 
the superpotential,
\be
{\cal W}=\sqrt{2}\sum_{i=1}^{N_f}\tilde{Q}_i(\Phi-m_i)Q_i
\nn\ee
Generically the masses break the flavour group of the theory 
$SU(N_f)\stackrel{m}{\longrightarrow} U(1)^{N_f-1}$. 
The Lagrangian also enjoys an $SU(2)_R\times U(1)_R$ classical R-symmetry. In 
the presence of non-zero masses, the latter is broken to $Z_2$. 

\para
The theory has an intricate moduli space of vacua depending on the hypermultiplet 
masses $m_i$, as well as a Fayet-Illiopoulos (FI) parameter which we shall 
introduce shortly. For now, we take this FI parameter to vanish, ensuring that 
there is always a 
Coulomb branch of vacua parameterised by $\phi={\rm diag}(\phi_1,\ldots,\phi_{N_{c}})$  
in which the gauge group is generically broken to the Cartan subalgebra 
$U(N_c)\stackrel{\phi}{\longrightarrow}U(1)^{N_c}$. 
When some of the masses coincide, one can also have Higgs branches of vacua 
parameterised by holomorphic gauge invariant operators formed from the hypermultiplet fields. 
For $N_f=N_c$, these include the baryonic operators, 
\be
B&=&Q_1^{a_1}Q_2^{a_2}\ldots Q_{N_c}^{a_{N_{c}}}\,\epsilon_{a_1\ldots a_{N_c}} \nn\\
\tilde{B}&=&\tilde{Q}_1^{a_1}\tilde{Q}_2^{a_2}\ldots \tilde{Q}_{N_c}^{a_{N_{c}}}
\,\epsilon_{a_1\ldots a_{N_c}}
\nn\ee
where $a_i$ denote colour indices. There are also meson operators of the form 
$M_{ij}=\tilde{Q}_iQ_j$. 

\para
The classical spectrum of BPS states depends on the vacuum in which the 
theory lives. We shall start by discussing the classical spectrum on the Coulomb branch, 
only subsequently moving onto quantum corrected spectrum and, ultimately, to the 
quantum spectrum on the 
Higgs branch. At a generic point on the Coulomb branch the theory has an interesting 
mixture of BPS states arising from both elementary excitations as well as 
non-perturbative monopole and dyon states. Among the former are the 
$N_c$ massless photons, together with $N_c(N_c-1)$ W-bosons with 
mass $|\phi_a-\phi_b|$ for $a,b=1\ldots, N_c$. There are also 
$N_cN_f$ BPS quark states which, for $a=1,\ldots,N_c$ and $i=1,\ldots N_f$ have 
masses given by,
\be
M_{\rm quark}=|\phi_a-m_i|
\label{quake}\ee 
All further BPS states arise as 
solitons and have non-zero magnetic charges under the unbroken gauge group 
$U(1)^{N_c}$. We denote these magnetic charges as $h_a$ and require $\sum_ah_a=0$, 
reflecting the fact that monopole solutions only exist in the semi-simple 
$SU(N)_C\subset U(N)_C$ 
part of the gauge group. The classical mass of these monopoles is given by 
\be
M_{\rm mon}=\frac{2\pi}{e^2}\left|\sum_{a=1}^{N_c}h_a\phi_a\right|
\label{monmass}\ee
In addition to these 
purely magnetic solitons, the classical spectrum also contains an infinite tower of 
dyons. A unified mass formula for each of these objects can be given in terms 
of the central charge $Z$. 
For BPS states with electric charge $j_a$ and magnetic charge $h_a$ under $U(1)^{N_c}$, 
and with charge $s_i$ under the global flavour group $U(1)^{N_f-1}$, the mass of 
any BPS state is given by $M=|Z|$ with
\be
Z=\sum_{a=1}^{N_c}\phi_a(j_a+\tau h_a)+\sum_{i=1}^{N_f}m_is_i
\label{z}\ee 
The above discussion has been classical. Let us now turn to various aspects 
of the quantum theory. The overall $U(1)$ part of the gauge group becomes weakly 
coupled in the infra-red\footnote{Readers uncomfortable with the Landau pole are 
free to turn on a noncommutivity parameter and repeat the story below.} and 
the interesting dynamics lies in the interactions of the $SU(N)$ part of the gauge group. 
For vanishing $m=\phi=0$, the one-loop beta-function for the $SU(N_c)$ gauge coupling has 
a coefficient 
proportional to $-(2N_c-N_f)=-N_c$ and the gauge coupling $e^2$ runs logarithmically 
with the scale $\mu$. It can be eliminated in favour of an RG invariant scale,
\be
\Lambda = \mu\exp\left(-\frac{4\pi^2}{N_c\,e^2(\mu)}\right)
\label{4dlm}\ee
Another quantum effect which 
will be important in the following arises from anomalies: the $U(1)_R$ symmetry 
is broken by instantons to ${\bf Z}_{2(2N_c-N_f)}={\bf Z}_{2N_c}$ when $m_i=0$. 
(Recall that, in the presence of hypermultiplet 
masses $m_i$, $U(1)_R$ is further broken at the classical level to $Z_2$).  

\para
Most important for our purposes are the quantum corrections to the masses 
of BPS 
states. At weak coupling $|\phi_a-\phi_b|\gg\Lambda$, one can show 
that the mass formula receives contributions from one-loop effects, together 
with an infinite series of instanton corrections. At strong coupling 
one needs another technique to compute the spectrum. Thankfully a 
beautiful method is provided by Seiberg and Witten's famous solution to the 
low-energy dynamics on the Coulomb branch \cite{sw}. We now review 
the Seiberg-Witten solution for the exact central charge $Z$ 
evaluated at a specific point on the Coulomb branch.

\subsection*{At the Root of the Baryonic Higgs Branch}

For reasons that will shortly become clear, we will be interested in 
the BPS spectrum of the theory arising at a point on the Coulomb 
branch known as the ``root of the baryonic Higgs branch''\footnote{In the 
present context, with $N_f=N_C$, there is no Higgs branch emanating from 
this point even when $m_i=0$. A better name might  be 
``root of the baryonic Higgs phase''.}  \cite{aps}.
This is the point defined classically by $\phi={\rm diag}(m_1,\ldots,m_{N_c})$. 
so that the breaking of flavour and gauge symmetries occurs at the same scale 
$U(N_c)\times SU(N_f) \stackrel{m}{\longrightarrow} U(1)^{N_c}\times U(1)^{N_f-1}$. 
From equation \eqn{quake} we see that $N_c$ of the $N_fN_c$ degrees of quark freedom 
become massless at this point. In fact, the quark masses become 
precisely degenerate with the masses of photons and W-bosons, each of 
which have classical masses for given by 
\be
M_{\rm W-boson}=M_{\rm quark} = |m_i-m_j|
\label{quarks}\ee
Because of 
this degeneracy the classical central charge \eqn{z} may be written in 
the simplified form,
\be
Z=\sum_{i=1}^{N_c}m_i(S_i+\tau h_i)
\label{zb}\ee
where we have redefined the charges as $S_i=s_a+j_a$. We would now like 
to describe the quantum corrections to this charge formula as encoded 
in the Seiberg-Witten solution. (Recently the semi-classical computation 
of corrections to the monopole mass was revisited in \cite{shif,peter}, finding  
agreement with the exact result of Seiberg and Witten).  
At the root of the baryonic Higgs branch, 
the Seiberg-Witten elliptic curve has a special property: it degenerates 
\cite{aps}
\be
F(t,u)=\left(t-\prod_{i=1}^{N_c}(u-m_i)\right)
\left(u-\Lambda^{N_c}\right)
\label{curve}\ee
This form of the curve occurs naturally in the 
M-theory construction of \cite{witm5}, where the degeneration corresponds 
to the fact that one of the IIA NS5 branes remains unbent upon its ascent 
to M-theory. The curve is branched over 
the $N_c$ points $e_{i}$ defined by, 
\be
\prod_{i=1}^{N_c}(u-m_{i}) -\Lambda^{N_c}= \prod_{i=1}^{N_c}(u-e_{i})=0
\label{corblimey}\ee
In the quantum theory the central charge is given by the integral 
of the Seiberg-Witten 
differential $\lambda_{SW}=(u/t)dt$ 
over certain one cycles of the curve. 
The resulting modification of the classical formula \eqn{zb} is 
\be
Z=\sum_{i=1}^{N_c}\, (m_{i}S_{i}+ m_{Di}h_{i}) 
\label{zend}\ee
where all the quantum corrections are encoded in the functions $m_{D\,i}$ which are 
holomorphic in 
the hypermultiplet masses $m_i$ and $\Lambda$. They are given by
\be
m_{Dl}-m_{Dk}=\frac{1}{2\pi i}\int_{e_{k}}^{e_{l}}d\lambda_{\rm SW}=
\frac{1}{2\pi i}\int_{e_{k}}^{e_{l}} u \frac{dt}{t} =  \frac{1}{2\pi}
\sum_{i=1}^{N_c}\,\int^{e_{l}}_{e_{k}} \frac{u\,du}{u-m_{i}} 
\nn\ee
where, in the final equality, we have used the exact form of the curve 
\eqn{curve}. Evaluating this integral, we find the expression for the 
contribution to the central charge given by
\be
m_{Dl}-m_{Dk} =  \frac{1}{2\pi} N_c(e_{l}-e_{k})+
\frac{1}{2\pi}\sum_{i=1}^{N_c}m_{i}\log\left(\frac{e_{l}-m_{i}}
{e_{k}-m_{i}}\right) 
\label{fr}\ee

\subsection*{On the Baryonic Higgs Branch}

The Seiberg-Witten computation of the spectrum holds on the Coulomb branch and 
we have presented the result above at a very specific point, known as the root 
of the baryonic Higgs branch. Let us now ask what becomes of the BPS spectrum as 
we move {\it onto} the baryonic Higgs branch. We do this by turning on a Fayet-Illiopoulos 
(FI) parameter $v^2$ for the $U(1)$ part of the gauge theory, so the D-term becomes,
\be
D=\sum_{i=1}^{N_f}q_iq^\dagger_i-\tilde{q}^\dagger_i\tilde{q}_i-v^2
\nn\ee
The FI parameter $v^2$ lifts the Coulomb branch and forces the theory onto the 
Higgs branch. The theory has a 
unique vacuum state, given by
\be
\phi={\rm diag}(m_1,\ldots, m_{N_c})\ \ \ \ ,\ \ \ \ \ B=v^{N_f}\ \ \ \ \ ,\ \ \ \ \ \  
\tilde{B}=M=0
\label{buffy}\ee
We now see why the point $\phi={\rm diag}(m_1,\ldots, m_{N_c})$ is called the 
root of the baryonic Higgs branch: it indeed provides the gateway into the 
Higgs phase when the FI parameter is turned on. The pattern of 
symmetry breaking in this vacuum is given by
\be
U(N_c)\times SU(N_f)
\stackrel{m}{\longrightarrow} U(1)^{N_c}\times U(1)^{N_f-1}
\stackrel{v}{\longrightarrow}U(1)^{N_c-1}_{\rm diag}
\label{break1}\ee
Our interest remains 
on the spectrum of BPS states, but now in the vacuum \eqn{buffy}. 
What becomes of the various BPS states as we turn on the FI parameter $v^2$? Let 
us firstly consider elementary excitations. The photons and W-bosons 
pick up an extra contribution to their mass 
proportional to $ev$ through the Higgs mechanism. In doing so, they combine with 
the $N_fN_c=N_c^2$ quark hypermultiplets and are no longer BPS, now sitting in long 
supersymmetry multiplets\footnote{This issue also arose in 
\cite{MQCD,decon} where it was argued that, in certain theories, they remain 
``almost BPS''.} \cite{vyung}. None of the elementary particle states 
remain BPS. 
\newcommand{\onefigurenocap}[1]{\begin{figure}[h]
         \begin{center}\leavevmode\epsfbox{#1.eps}\end{center}
         \end{figure}}
\newcommand{\onefigure}[2]{\begin{figure}[htbp]
         \begin{center}\leavevmode\epsfbox{#1.eps}\end{center}
         \caption{\small #2\label{#1}}
         \end{figure}}
\begin{figure}[htbp]
\begin{center}
\epsfxsize=4.5in\leavevmode\epsfbox{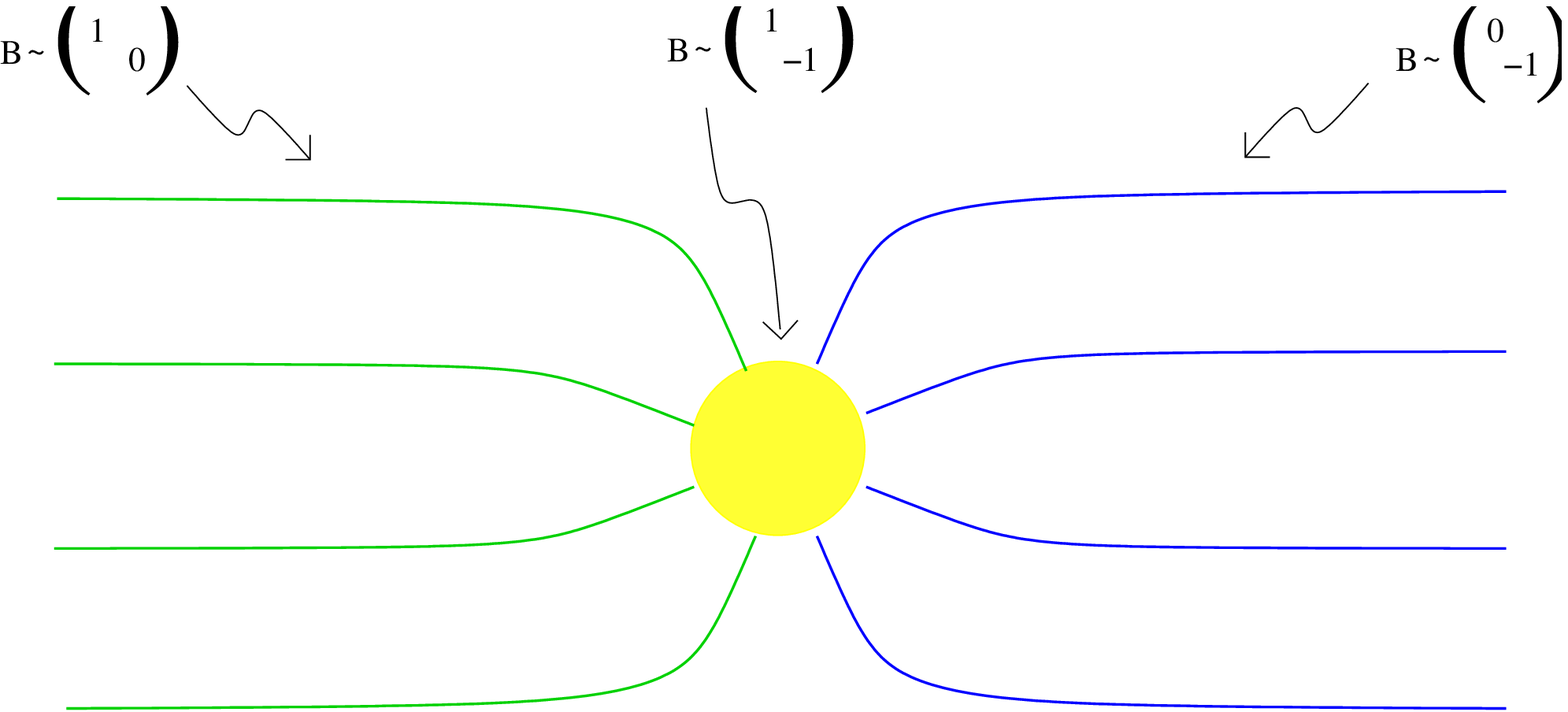}
\end{center}
 {\small Figure 1: A sketch of the $U(2)$ monopole in the Higgs phase \cite{monflux}. 
The caricature is accurate for $ev\ll |m_i-m_j|$, a limit which ensures the flux tube 
is larger than the monopole core.}
\end{figure}
\para
Let us now turn to the magnetic monopoles. At first sight, it appears unlikely 
that they can remain BPS on the Higgs branch. Since the gauge group is fully 
broken, the Meissner effect ensures that magnetic flux can no longer freely 
permeate the vacuum but is restricted to lie in a flux tube. Thus the monopoles are 
confined and, in isolation, have infinite mass. Nevertheless, as shown in \cite{monflux}, 
the monopoles are BPS. The final object can be thought of 
as the original monopole, now emitting two vortex strings and the total combination 
preserves $1/4$ of the original supersymmetry. 
The classical Bogomoln'yi equations describing this monopole-flux-tube combo 
can be derived by completing the square in the Hamiltonian. 
Denoting the non-abelian magnetic 
field as $B$ and setting all irrelevant fields to zero we choose that the 
monopole ejects its flux tubes in the $x^3$ directions. We manipulate 
the Hamiltonian thus \cite{monflux}
\be
{\cal H}&=&\frac{1}{2e^2}B^2+\frac{1}{2e^2}|{\cal D}\phi|^2+|{\cal D}q_i|^2 
+\frac{e^2}{2}(q_iq_i^\dagger-v^2)^2+q_i^\dagger|\phi-m_i|^2q_i \nn\\
&=& 
\frac{1}{2e^2}({\cal D}_1\phi-B_1)^2 + \frac{1}{2e^2}({\cal D}_2\phi-B_2)^2 
+({\cal D}_3\phi-B_3-e^2(q_iq_i^\dagger-v^2))^2 \nn\\ && 
+|{\cal D}_1q_i-i{\cal D}_2q_i|^2+|{\cal D}_3q_i+(\phi-m_i)q_i|^2 
+\frac{1}{e^2}\partial_\mu(\phi B_\mu) -v^2B_3\nn\\ &\geq&
\frac{1}{e^2}\partial_\mu(\phi B_\mu)-v^2B_3
\label{ham}\ee
where we have left colour indices and traces implicit and we have summed over the flavour 
index $i$. The Bogomoln'yi equations can be found in the total squares on 
the second line. While no explicit solutions to these equations are known, several 
properties were deduced in \cite{monflux}\footnote{As explained in \cite{shif}, 
solutions to these equations describe  a wider range of objects than the confined monopoles 
considered here and include strings ending on domain walls.}. We draw a caricature of the solution 
in the Figure.

\para
The two terms in the final line of \eqn{ham} measure 
conserved topological charges. The first is precisely the magnetic charge carried 
by the monopole. In 
the Coulomb phase the integral $\int d^3x \ \partial\cdot(\phi B)$ is evaluated 
on the $S^2_\infty$ boundary. In the present case, the monopole flux does 
not make it to all points on the boundary, but is confined to two 
flux tubes which stretch in the $\pm x^3$ direction. Correspondingly, the 
integral should now be evaluated over two planes ${\bf R}^2_{\pm\infty}$ at 
$x^3=\pm\infty$. The second term in \eqn{ham} is new. When integrated over 
the $x^1-x^2$ plane, it measures the tension of the flux tubes emitted by the 
monopole. These are simply vortices, supported by the overall broken $U(1)$ 
gauge symmetry in \eqn{break1}. They have tension given by 
$T_{\rm vortex}=2\pi v^2 k$ for $k\in {\bf Z}$ and will be the subject of study in 
Section 3. Since these vortex strings have finite tension 
and semi-infinite length, the total mass of the configuration is 
infinite. This reflects the fact that the monopoles are confined. Nonetheless, 
as we see from \eqn{ham}, this infinite mass splits unambiguously 
into a finite contribution from the monopole and an infinite contribution from the 
flux-tube. With the monopole's mass defined in this way, we see that it 
remains identical to that calculated in the Coulomb phase \eqn{monmass}.

\para
In summary, on the Higgs branch the quarks and W-bosons combine to form 
long multiplets, while the monopoles are confined yet remain BPS. 
Moreover, after subtracting the contribution from the BPS flux tube, 
we have seen that the classical monopole mass remains unchanged as we turn on 
the FI parameter $v^2$ and move onto the Higgs branch. Can we understand this and 
extend the result to the quantum theory? 
In fact, there is a simple non-renormalisation 
theorem that tells us that the central charge $Z$ for particle states 
cannot receive contributions from the FI parameter $v^2$ and remains given by 
\eqn{zend} for BPS states on the Higgs branch. 
The important observation is the fact that, in 
${\cal N}=2$ theories, the central charges are given by the scalar components of 
background vector multiplets \cite{sw}. Any dependence on hypermultiplets or linear 
multiplets (also known as tensor multiplets) is forbidden by supersymmetry. 
The FI parameter $v^2$ lies in a background linear multiplet (it is actually one  
component of a triplet of FI parameters which is precisely the scalar field content 
of the ${\cal N}=2$ linear multiplet). We therefore conclude that the BPS particle 
states receive no contribution to their mass from $v^2$ and the exact quantum 
corrected central charge on the Higgs branch is given by \eqn{zend}.

\section{The Vortex Theory}

In the previous Section we have derived the BPS spectrum on the baryonic Higgs 
branch. We have seen that there are no vector multiplet BPS states, but quarks 
and monopoles both survive as BPS objects. While the quarks interact only through 
short range forces, the monopoles are confined by the Meissner effect. Moreover, we have 
something new: a BPS vortex string with tension $2\pi v^2$. In this Section, 
we study the quantum dynamics of this vortex string and show that its mass spectrum 
reproduces the four dimensional BPS spectrum described above.

\para
Let us start by describing the vortex in the theory with vanishing quark masses, $m_i=0$. 
In this case, the Lagrangian preserves the full $SU(N_f)$ flavour symmetry but the unique 
vacuum state on the Higgs branch lies in a colour-flavour locked phase with the 
symmetry breaking pattern $U(N_c)\times SU(N_f) \stackrel{v}{\longrightarrow} SU(N_c)_{\rm diag}$. 
The breaking of the overall $U(1)$ gauge group ensures that vortex strings are supported 
with topological winding number given by $\int {\rm Tr}\, B = 2\pi k$, with $k\in {\bf Z}$, where 
the integral is taken over the plane transverse to the string. If we choose the strings to lie 
in the $x^3$ direction, then the classical configurations obey the non-abelian version of the 
first order vortex equations
\be
B_3= e^2(\sum_{i=1}^{N_f}q_iq_i^\dagger - v^2)\ \ \ \ \ ,\ \ \ \ \ {\cal D}_1q_i=i{\cal D}_2q_i
\label{harry}\ee
The strings are BPS with tension given by $T_{\rm vortex}= 2\pi v^2 k$. In \cite{vib}, an 
ADHM-like construction of the $k$-vortex moduli space was derived from a D-brane picture. 
We review this in Appendix A. 
In the remainder of this paper we shall content ourselves with studying a single vortex $k=1$. 
In this case, all zero modes of the vortex are Goldstone modes and the moduli space can 
be constructed simply from the symmetries of the field theory \cite{vib,yung}. 
The key point to note is that a single non-abelian vortex is simply an abelian 
Nielsen-Olesen vortex embedded into a $U(1)$ subgroup of $U(N_c)$. Suppose we choose 
to embed it in the upper left-hand corner. Then acting on this solution with the 
$SU(N_c)_{\rm diag}$ vacuum symmetry sweeps out a moduli space 
${\cal M}_{\rm vortex}\cong SU(N_c)/(U(1)\times SU(N_c-1))\cong {\bf CP}^{N_c-1}$ 
of solutions. We therefore have,
\be
{\cal M}_{\rm vortex} = {\bf C}\times {\bf CP}^{N_c-1}
\label{moduli}\ee
where ${\bf C}$ parameterises the center of mass of the vortex string 
in the $x^1-x^2$ plane, while ${\bf CP}^{N_c-1}$ describes the internal 
degrees of freedom arising from the $SU(N_c)_{\rm diag}$ action. The low-energy 
dynamics of the vortex string can be described by a $d=1+1$ dimensional 
sigma-model with target space ${\cal M}_{\rm vortex}$. Since the vortex is 
BPS, the low-energy dynamics preserves ${\cal N}=(2,2)$ supersymmetry.

\para 
Let us ask how this situation changes for non-zero quark masses $m_i$. The answer 
was given in \cite{monflux}. The masses break the symmetry group as 
$SU(N_c)_{\rm diag}\rightarrow U(1)^{N_c-1}_{\rm diag}$, 
lifting the ${\bf CP}^{N_c-1}$ moduli space. For a vortex of unit winding 
number, there are now $N_c$ isolated solutions corresponding to an abelian vortex 
embedded in one of the {\it diagonal} $U(1)\in U(N_c)$ subgroups. In other words, 
the off-diagonal embeddings have been removed. From the perspective 
of the low-energy dynamics, the masses $m_i$ induce a potential $V$ on ${\bf CP}^{N_c-1}$ 
with $N_c$ isolated minima. This potential is of the form $V\sim K^2$ where 
$K$ is a holomorphic Killing vector on ${\bf CP}^{N_c-1}$. We derive this potential 
in Appendix B.

\para
We now describe the theory in more detail and flesh out some of these results. 
Firstly, note that our low-energy approach to determine the spectrum of the 
string is a priori trustworthy provided the string is sufficiently massive: $ev \gg |m_i-m_j|$. 
In fact, because of the BPS nature of our results, 
they can ultimately be continued throughout parameter space. With this in mind, 
we now describe the theory of the vortex. We use the language of the 
gauged-linear sigma model. This description arises naturally in the brane picture of 
\cite{vib} which we review in Appendix A

\para
{\bf Vortex Theory:} {\it $d=1+1$, ${\cal N}=(2,2)$ supersymmetric $U(1)$ with a single
neutral chiral multiplet $Z$ and $N_c$ chiral multiplets $\Psi_i$ of charge $+1$. Each charged 
chiral has twisted masses $m_i$, $i=1,\dots, N_c$. 
The classical theory has dimensionless FI parameter $r$ and vacuum angle $\theta$ 
which are combined in a single complex coupling $\tau=ir+\theta/2\pi$. 
The gauge theory also contains a dimensionful gauge coupling $g$. }

\para
A couple of comments are in order. Firstly, the twisted mass in two dimensional 
gauge theories was introduced in \cite{hh}. Each twisted mass is a complex mass for a 
chiral multiplet, consistent with supersymmetry and gauge invariance. It is forbidden 
in four-dimensional ${\cal N}=1$ theories by Lorentz symmetry, but becomes available upon 
dimensional reduction to two dimensions. As our notation suggests, the 
twisted masses of the vortex theory are identified with the hypermultiplet masses $m_i$  
in four-dimensions. This follows immediately from the brane picture of \cite{hh} and \cite{vib}. 
The FI parameter of the vortex theory, which determines the 
K\"ahler class  of the ${\bf CP}^{N_c-1}$ moduli space can also be extracted 
from the brane construction \cite{vib}
\be
r=\frac{2\pi}{e^2}
\label{lovely}\ee
Note that, with this result, the complexified coupling $\tau$ of the vortex theory 
is identified with the complexified coupling $\tau$ of the four-dimensional 
theory\footnote{The identification of the theta angle in four dimensions with the 
theta angle on the vortex theory is new. It follows simply from the IIA version of 
the brane construction in \cite{vib}. For both the two dimensional theory on 
the vortex \cite{hh} and the four dimensional theory \cite{witm5}, the theta angle is 
given by the separation of M5-branes along the M-theory circle.}.
Finally, we are 
instructed in \cite{vib} to take the two-dimensional gauge coupling  
$g^2\rightarrow \infty$. This arises as a consequence of the decoupling limit of 
the D-brane system and forces the vortex theory onto its Higgs branch. In what follows, we 
will leave $g^2$ finite. This is justified by the existence of the CFIV supersymmetric 
index \cite{cfiv} which ensures that the BPS spectrum of the vortex theory is independent of $g^2$.

\para
The neutral chiral multiplet $Z$ contains a single complex scalar field $z$, 
parameterising the center of mass motion of the vortex. It corresponds to the 
${\bf C}$ factor in \eqn{moduli}. Since this field is free, we pay it no more 
attention and ignore it in the following.  Each charged chiral multiplet $\Psi_i$ also 
contains a complex scalar $\psi_i$, $i=1,\ldots,N_c$, while the $U(1)$ vector multiplet 
contains the two dimensional gauge field and a further, neutral, complex scalar $\sigma$. 
The bosonic part of the Lagrangian describing the internal degrees of freedom of 
the vortex is given by,
\be
-{\cal L}_{\rm vortex} =  \frac{1}{2g^2}\left(F_{01}^2+|\partial\sigma|^2\right) 
+\sum_{i=1}^{N_c}\left(|{\cal D}\psi_i|^2 + |\sigma-m_i|^2|\psi_i|^2
\right)+\frac{g^2}{2}(\sum_{i=1}^{N_c}|\psi_i|^2-r)^2
\label{ours}\ee
For vanishing twisted masses $m_i$, the theory has a $SU(N_c)_D$ global symmetry 
which is identified with the $SU(N_c)_{\rm diag}$ symmetry in four dimensions. For 
generic $m_i\neq 0$, this is broken to $U(1)^{N_c-1}_D$. The theory also has a $U(1)_R$ 
symmetry which is inherited from the $U(1)_R$ symmetry in four dimensions. This 
rotates the phases of both $\sigma$ and $m_i$. For vanishing masses, the 
vortex theory has a Higgs branch of vacua given by $\sigma=0$ with the chiral 
multiplets constrained to obey $\sum_i|\psi|^2=r$. After dividing by the $U(1)$ 
action we see the Higgs branch is ${\bf CP}^{N_c-1}$ in agreement with \eqn{moduli}. 
In the presence of twisted masses, performing the same procedure results in a  
twisted potential on the Higgs branch of the type constructed in \cite{agf} as we 
show explicitly in Appendix B. The potential has $N_c$ isolated vacua given by,
\be
{\rm Vacuum}\ i: \ \ \ \sigma=m_i\ \ \ \ \ ,\ \ \ \ \ |\psi_j|^2=r\delta_{ij}
\label{vac}\ee 
As described above, the  
i${}^{\rm th}$ vacuum corresponds to a vortex embedded in the i${}^{\rm th}$ 
$U(1)$ subgroup, carrying magnetic charge $B={\rm diag}(0,\ldots,0,1,0,\ldots, 0)$, 
where the $1$ sits in the i${}^{\rm th}$ entry.

\para
So far we have discussed the relevant aspects of the classical two-dimensional 
theory on the vortex worldsheet. Let us now turn to the quantum theory. 
When the twisted masses vanish $m_i=0$, 
there is a one-loop correction to the FI parameter $r$, leading to 
a logarithmic running at scale $\mu$,
\be
r(\mu) = r_0 -\frac{N_c}{2\pi}\log\left(\frac{M_{UV}}{\mu}\right)
\label{rmu}\ee
where $r_0$ is the bare FI parameter defined at the UV cut-off $M_{UV}$. 
Note that, since this theory describes the low-energy dynamics of a 
soliton, it is inappropriate to take $M_{UV}$ to infinity. Instead it 
is set by the mass scale of the vortex: $M_{UV}=v^2$. 

\para
In \eqn{rmu} we see our first hint that the vortex theory understands something of 
the four dimensional quantum dynamics since the one-loop beta function 
for $r$ is identical to that of the four-dimensional coupling $e^2$. This 
ensures that the relationship 
$r=2\pi/e^2$ is preserved under RG flow. Note that although vortices 
exist by virtue of the overall $U(1)\subset U(N_c)$, the renormalisation 
of $r$ clearly follows the asymptotically free $SU(N_c)$ gauge coupling in 
four dimensions, rather than the infra-red free $U(1)$ coupling. 
Since the beta functions for $r$ and $2\pi/e^2$ are equal, it follows 
that if we eliminate $r(\mu)$ in favour of the one-loop RG invariant scale,
\be
\Lambda=\mu\exp\left(-\frac{2\pi r(\mu)}{N_c}\right)
\nn\ee
then this coincides with the dynamically generated scale in four dimensions 
\eqn{4dlm}. 

\para
The anomaly structure provides further agreement between the vortex theory 
and four dimensions. The $U(1)_R$ symmetry on the vortex worldsheet is 
broken by anomalies to ${\bf Z}_{2N_c}$, in agreement with the four dimensional 
result. This suggests an interplay between Yang-Mills instantons and worldsheet 
instantons. We shall return to this later.

\para 
In the presence of twisted masses, the story is similar. The running
of the coupling $r(\mu)$ is cut-off at the scale $|m_i-m_j|$. For 
$|m_i-m_j|\gg \Lambda$, the theory is weakly coupled. Again, this 
is in agreement with the four dimensional theory at the root of 
the baryonic Higgs branch, which sits far out on the Coulomb branch when 
$|m_i-m_j|\gg \Lambda$.  
In this regime, the $N_c$ classical vacua of the vortex theory \eqn{vac} 
are trustworthy ground states around which to study excitations. 
Finally, we note that at strong coupling, $|m_i-m_j|\ll \Lambda$, the 
Witten index ensures that there remain $N_c$ isolated vacuum states in 
the quantum vortex theory.

\subsection*{The Spectrum of the Vortex String}

Having identified the theory on the vortex string and described some 
of its properties, our task now is to determine its spectrum. In 
fact this is precisely the calculation performed by Dorey in \cite{nick} 
where he computed the exact quantum BPS spectrum as a function of 
the twisted masses $m_i$ and $\Lambda$. In this subsection we review the results 
of \cite{nick} and describe how they relate to the vortex string.

\para
We deal first with the classical, elementary internal excitations of the BPS string. 
The vortex theory \eqn{ours} includes a gapped photon with mass $g\sqrt{r}$. 
This does not lie in a BPS multiplet and, moreover, decouples as 
$g^2\rightarrow\infty$ so we do not consider it in the following. The elementary 
BPS excitations arise from the chiral multiplets $\psi_i$. As we 
have seen, when the quark masses $m_i$ vanish these parameterise the massless 
Goldstone modes of the internal ${\bf CP}^{N_c-1}$ vortex moduli space of \eqn{moduli}. In the 
presence of the masses $m_i$, these flat directions are
lifted and the vortex theory has a classical 
mass gap. In the $i^{\rm th}$ vacuum, there are $(N_c-1)$ BPS states arising 
from the $\psi_j$ with, for $j\neq i$,  masses given by 
\be
M_{\psi} = |m_j-m_i|
\label{nonny}\ee
We see that these perturbative excitations of the string reproduce the classical mass spectrum of 
the quarks and W-bosons \eqn{quarks} in the four dimensional theory, but on the Coulomb branch. 
Recall that, in the Higgs vacuum we are considering, the classical mass of these 
particles is increased by a contribution from $ev$ and they are no longer BPS. 
How then can we understand the agreement of the BPS formula \eqn{nonny} on the string and 
four dimensional BPS formula on the Coulomb branch \eqn{quarks}? 
These elementary states of the vortex are to be thought of as four-dimensional elementary 
particles bound to the string, an interpretation which is clear from the brane picture of 
Appendix A. In the center of the vortex string, one of the Higgs fields $q_i$ vanishes and the 
theory effectively sits in a partial Coulomb phase\footnote{We thank M. Shifman and 
A. Yung for discussions and suggestions on this point.}. 
The W-bosons and quarks which are carry charge under the 
corresponding $U(1)$ may lower their mass to their Coulomb branch value by sitting where 
$q_i=0$. For the $i^{\rm th}$ vortex, these are precisely the states with mass \eqn{nonny}. 
The calculation above shows that these states actually re-obtain 
BPS status by this mechanism.

\para 
When the classical vortex theory has isolated vacua, it also admits topological 
kink solutions which contribute to the spectrum. Kinks in models of 
this type have been much studied in the literature, starting with Abraham 
and Townsend \cite{at} and continued in \cite{nick,kinky,more}. 
The first order Bogomoln'yi equations describing the kink are given by,
\be
\partial\sigma &=& g^2(\sum_{i=1}^{N_c}|\psi_i|^2-r) \nn\\
{\cal D}\psi_i&=&(\sigma-m_i)\psi_i
\nn\ee
where all derivatives are along the spatial worldvolume direction of the 
vortex string, and the fields are 
subject to the boundary conditions that they return to Vacuum $i$ as $x\rightarrow -\infty$, 
and to Vacuum $j$ as $x\rightarrow +\infty$. The BPS mass of such a kink is given by,
\be
{\cal M}_{\rm kink} = r|m_i-m_j|=\frac{2\pi}{e^2}|m_i-m_j|
\label{mkink}\ee
Comparing with equation \eqn{monmass}, we see that this coincides with the 
mass of the monopole with magnetic charge $h_a=\delta_{ai}-\delta_{aj}$, 
sitting at the root of the baryonic Higgs branch. In fact, as shown in 
\cite{monflux}, the kink in the vortex string is precisely this magnetic monopole 
in the Higgs phase, with the string providing the flux line which whisks 
away the magnetic charge as required by Meissner. 
To see this, we examine the quantum numbers of the kink. 
As $x\rightarrow -\infty$, the vortex theory sits in the i${}^{\rm th}$ vacuum 
state, corresponding to a magnetic flux in $U(1)_i\subset U(1)^{N_c}$. In the other 
direction, as $x\rightarrow +\infty$, the vortex sits in the j${}^{\rm th}$ vacuum, 
the magnetic flux in the $U(1)_j\subset U(1)^{N_c}$ subgroup. 
Taking into account the direction of the flux, we see that the kink must 
provide a source of magnetic charge  $h_a=\delta_{ai}-\delta_{aj}$, 
precisely that of the monopole. The magnetic flux assignment for a $U(2)$ monopole is
drawn in the Figure.

\para 
Finally, as with the monopoles of Section 2, the kinks on the vortex string also 
admit a generalisation to dyons in which they are charged under the 
$U(1)^{N_c-1}_{D}$ global flavour group of the vortex theory \cite{at}. Such 
objects are known as q-kinks. Moreover, 
there is also an analog of the Witten effect \cite{witteff} for these kinks so 
that, in the presence of a $\theta$-angle, they pick up global electric charge \cite{nick}. 

\para
To summarise, the classical BPS spectrum on the vortex string 
consists of a rich mix of both elementary and topological excitations. To write a 
central charge formula for the masses, we define the charge of a state under the 
$U(1)^{N_c-1}_{D}$ global flavour symmetry to be $S_i$. We further define the 
topological charge $T_i$, such that a field configuration that tends toward 
Vacuum $j$ as $x\rightarrow -\infty$ and to Vacuum $k$ as $x\rightarrow +\infty$ 
has topological charge $T_i=\delta_{ij}-\delta_{jk}$. The masses of all 
BPS states are then given by $M=|Z|$ with the classical central charge given by,
\be
Z=i\sum_{i=1}^{N_c}m_i(S_i+\tau T_i)
\nn\ee
which agrees precisely with the classical central charge of the four-dimensional 
theory \eqn{zb} if we equate the two-dimensional topological charge with the 
four-dimensional magnetic charge: $T_i=h_i$. 

\para
Now we turn to the description of the quantum spectrum of the vortex string. Once 
again exact results are available, although of a very different nature from the 
Seiberg-Witten curve that we employed in Section 2. The 
trick, following Witten \cite{glsm}, is to integrate out the chiral superfields 
$\Psi_i$ leaving an effective Lagrangian for the vector multiplet fields. 
This is most elegantly expressed in terms of a twisted chiral superfield 
$\Sigma$ whose lowest component is the complex scalar field $\sigma$, and 
includes $F_{01}$ as part of the auxiliary field. In the presence of 
twisted masses, this calculation was first done in \cite{hh}, resulting in the effective 
twisted superpotential,
\be
{\cal W}(\Sigma) = \frac{i}{2}\tau\Sigma - \frac{1}{4\pi}\sum_{i=1}^{N_c}
(\Sigma-m_i)\log\left(\frac{2}{\mu}(\Sigma-m_i)\right)
\nn\ee
Assuming no singularities in the K\"ahler potential, the $N_c$ quantum vacua 
of the theory are determined by the critical points of the twisted superpotential 
$\partial{\cal W}/\partial\Sigma = 0$ and are given by,
\be
\prod_{i=1}^{N_c}(\sigma-m_i)-\Lambda^{N_c}\equiv \prod_{i=1}^{N_c}(\sigma - e_i)=0
\nn\ee
which we notice as the same equation describing the branch points of the 
Seiberg-Witten curve at the root of the baryonic Higgs branch \eqn{corblimey}. 
The classical BPS kinks which we described above also survive in this effective 
theory \cite{cv} although their mass is now corrected to include quantum effects. 
A kink interpolating between the Vacuum $i$ and Vacuum $j$ has mass 
$M_{\rm kink}=2\Delta{\cal W}=2{\cal W}(e_i)-2{\cal W}(e_j)$. In the weak 
coupling regime $|m_i-m_j|\gg\Lambda$ the leading contribution is precisely 
the classical result \eqn{mkink}. Deep in the strong coupling regime, 
$|m_i-m_j|\ll \Lambda$, quantum effects are dominant. The exact BPS mass of 
the kink can be captured by a correction to the central charge so that all 
BPS excitations of the string have masses $M=|Z|$, now with
\be
Z=-i\sum_{i=1}^{N_c}(m_iS_i+m_{D\,i}T_i)
\nn\ee
where all the quantum corrections are encoded in $m_{D, i}$, each a 
holomorphic function of $m_j$ and $\Lambda$. Using the expressions above, 
we find that (up to an $i$-independent irrelevant constant)
\be
m_{D\, i} = -2i{\cal W}(e_i) = \frac{1}{2\pi i}N_ce_i + \frac{1}{2\pi i}
\sum_{j=1}^{N_c}m_j\log \left(\frac{e_i-m_j}{\Lambda}\right)
\nn\ee
which we see coincides with the expression computed in four dimensions \eqn{fr}. 
Note that these two equations arose from very different origins: the degeneration 
of the Seiberg-Witten elliptic curve in four dimensions, and the critical 
points of the effective twisted superpotential in two dimensions. This agreement 
is the main result of \cite{nick}.

\para
Note that while we have shown, following \cite{nick}, that the exact central 
charges agree in two and four dimensions, this does not necessarily imply that 
the spectra coincide. For this we have to show that the same quantum numbers 
$S_i$ and $T_i$ are realised in each theory. For example, from the perspective 
of the vortex string, we have seen that only kinks with quantum numbers 
$T_i=\delta_{ij}-\delta_{ik}$ are allowed classically. In contrast, in the 
four dimensional theory, there exist classical monopole configurations with 
arbitrary magnetic charge $T_i$, subject only to $\sum_{i}T_i=0$. However, not 
all of these classical configurations may be realised as states in the 
quantum theory. It was shown in \cite{nick} that at weak coupling $|m_i-m_j|\gg \Lambda$, the 
allowed charges of quantum states do coincide between the two theories. Moreover, 
since the central charges agree, the curves of marginal stability where states 
may decay also coincide in the two theories. This strongly suggests that the 
spectra agree throughout the parameter space.

\subsection*{A Weak Coupling Expansion}

The results of the previous section reveal that the exact BPS mass spectrum of the 
vortex theory coincides with the exact BPS mass spectrum of the four-dimensional 
gauge theory. Powerful as these results are, it is constructive to examine them 
in the weak-coupling regime $|m_i-m_j|\gg \Lambda$. In this case, each 
holomorphic function $m_{D\,i}$ has the expansion, 
\be
m_{D\,i}=\frac{1}{2\pi i}\left(N_cm_i+\sum_{j=1}^{N_c}(m_j-m_i)\log\left(
\frac{m_j-m_i}{\Lambda}\right)+\sum_{n=1}^\infty c_n(m_j)\Lambda^n\right)
\nn\ee
where the $\log$ term arises as a one-loop contribution, while each term 
in the sum is due to a charge $n$ instanton effect with an $m_j$ dependent 
coefficient $c_n$. In the four-dimensional 
theory these are $U(N_c)$ Yang-Mills instantons while, in the theory 
on the vortex string worldsheet, they are two-dimensional instantons 
which are usually referred to as semi-local vortices or ${\bf CP}^{N_c-1}$ 
lumps. In other words, from the perspective of the vortex string, Yang-Mills 
instantons look like semi-local vortices: a vortex within a 
vortex\footnote{From brane picture this is clear. In the IIA 
T-dual version of the set-up in \cite{vib}, 
both objects arise as Euclidean D0-branes lying in the D4 world-volume.}. 
This is entirely analogous to 
the fact that, as we have seen above, a Yang-Mills monopole looks like a kink 
within a vortex. 
The fact that the coefficients $c_n$ coincide term by term is presumably related to the 
observation of \cite{vib} that the moduli space of semi-local vortices is a submanifold of 
the moduli space of Yang-Mills instantons. It would be interesting to understand this 
agreement at the semi-classical level.

\para
In fact, just as we derived the $1/4$-BPS 
Bogomoln'yi equations for the monopole in the vortex \cite{monflux}, 
we may similarly derive the equations describing the Yang-Mills instanton 
in the Higgs phase in the presence of the vortex string. 
To do so, we set the hypermultiplet 
masses $m_i=0$ to zero and work in four-dimensional Euclidean space. 
We define a complex structure on ${\bf R}^4$ given by $z=x^2+ix^3$ and 
$w=x^4+ix^1$, and complete the four-dimensional action thus,
\be
{\cal L}&=&\frac{1}{2e^2}F_{\mu\nu}F^{\mu\nu}+\sum_{i=1}^{N_f}|{\cal D}_\mu q_i|^2 
+\frac{e^2}{2}(\sum_{i=1}^{N_f}q_iq_i^\dagger - v^2)^2 \nn\\
&=& \frac{1}{2e^2}\left(F_{12}-F_{34}-e^2(\sum q_iq_i^\dagger-v^2)\right)^2 
+\sum\left(|{\cal D}_zq_i|^2+{\cal D}_{\bar{w}}q_i|^2\right) \nn\\ && 
+\frac{1}{2e^2}(F_{14}-F_{23})^2 +\frac{1}{2e^2}(F_{13}+F_{24})^2 
+\frac{1}{e^2}F_{\mu\nu}F_{\rho\sigma}\epsilon^{\mu\nu\rho\sigma}
+F_{12}v^2 - F_{34}v^2
\nn\\
&\geq &\frac{1}{e^2}F_{\mu\nu}F_{\rho\sigma}\epsilon^{\mu\nu\rho\sigma}
+F_{12}v^2 - F_{34}v^2
\nn\ee
The terms left in the final line are all topological charges. We recognise 
the first as counting instanton number $n$ when integrated over ${\bf R}^4$. 
The remaining two charges both count vortex strings. The term $F_{12}$ 
is the topological charge for a string extended in the $x^3-x^4$ plane 
as we have discussed above. The presence of the third charge $F_{34}$, 
which counts strings with worldvolume $x^1-x^2$, reflects the fact that 
the most general solution to the Bogomoln'yi equations appear to contain 
more than we bargained for: orthogonal vortex strings, which share no worldvolume directions, 
together with Yang-Mills instantons. The Bogomoln'yi equations are
\be
F_{12}-F_{34}=e^2(\sum_iq_iq_i^\dagger -v^2)\ \ ,\ \  
F_{14}=F_{23}\ \ , \ \ F_{13}=F_{24}\ \ ,\ \ {\cal D}_zq_i=0\ \ , \ \ 
{\cal D}_{\bar{w}}q_i=0
\nn\ee
We see that these are an interesting mix of the usual self-dual Yang-Mills equations 
and the non-abelian vortex equations \eqn{harry}. As we mentioned, the most general 
solution seems likely to describe 
$k_1$ vortices with worldvolume in the $x^3-x^4$, another $k_2$ vortices 
with worldvolume $x^1-x^2$, and $n$ Yang-Mills instantons. Such solutions 
likely preserve $1/8$ supersymmetry. It would be 
interesting to study the properties of these solutions further. The relevance for 
the current work is restricted to the $1/4$-BPS configurations with $k_2=0$.

\section{$N_f>N_c$ and Semi-Local Vortices}

In this Section we would like to generalise the story to ${\cal N}=2$ 
$U(N_c)$ supersymmetric QCD with $N_f > N_c$ massive hypermultiplets. 
In \cite{dht} a two-dimensional, non-compact sigma-model was presented 
whose mass spectrum coincides with that of this four-dimensional 
gauge theory. Here we confirm, using the results of \cite{vib}, that 
this is indeed the theory living on the vortex string. 

\para
We start by taking generic masses for the hypermultiplets $m_i\neq m_j$ 
for $i\neq j$ where $i$ and $j$ now run from $1$ to $N_f>N_c$. We also 
include a FI parameter $v^2$ from the beginning. The theory now has 
$N_f!/N_c!(N_f-N_c)!$ isolated vacua, labeled by the choice of $N_c$ 
quarks which have an expectation value. Without loss of generality, we 
may choose the vacuum $\phi={\rm diag}(m_1,\ldots,m_{N_c})$ with 
$\tilde{q}_i=0$ and 
\be
q^a_{\ i}=\left\{\begin{array}{ll} v\delta^a_{\ i}\ \ \ \ \ \ \ \ \  & a,i=1,\ldots, N_c \nn\\
0 & i=N_c+1,\ldots,N_f\end{array}\right.
\nn\ee
We will be interested in the BPS spectrum at this point on the Higgs branch. As in 
Section 2, the W-bosons combine with $N_c^2$ quarks to form long multiplets. 
However, in contrast to the theory with $N_f=N_c$, there are now  
BPS quark states. These arise from the $(N_f-N_c)N_c$ quark hypermultiplet which 
parameterise the flat directions of the Higgs branch when $m_i=0$. For non-vanishing 
$m_i$, these BPS quarks states have mass $M_{\rm quark} = |m_i-m_j|$ 
for $i=1,\ldots, N_c$ and for $j=N_c+1,\ldots, N_f$. The classical monopole 
spectrum remains much as in Section 2. We can again compute the 
quantum corrections to the central charge using the Seiberg-Witten curve at the 
corresponding point 
on the Coulomb branch. The relevant formulae can be found in \cite{dht} so 
we shall be brief: the classical central charge at a generic point on the 
Coulomb branch is again given by \eqn{z}. In our vacuum of choice  
$\phi={\rm diag}(m_1,\ldots,m_{N_c})$ the degeneracies in the spectrum between  
quarks and W-bosons allows us to simplify the central 
charge as
\be
Z=\sum_{i=1}^{N_c}m_i(S_i+m_{D\,i}h_i)+\sum_{i=N_c+1}^{N_f}m_is_i
\label{newz}\ee
where the definitions are as in Section 2, equation \eqn{zb} and, classically, 
$m_{D\,i}=\tau$. Quantum mechanically, $m_{D\,i}$ can again be expressed 
as the integral of the Seiberg-Witten one-form $\lambda_{SW}$ over a particular 
one-cycle of a new elliptic curve with $N_c$ branch points $e_i$. Skipping the 
details, we simply quote the final result: $m_{D\,i}$ takes the form,
\be
m_{D\,j} =  \frac{1}{2\pi}\left[(2N_c-N_f)e_{j} - 
\sum_{i=1}^{N_c}m_{i}\log\left(\frac{e_{j}-m_{i}}{\Lambda}\right) 
+\sum_{i=N_c+1}^{N_f}m_i\log\left(\frac{e_j-m_i}{\Lambda}\right)\right]
\label{doogiehowsermd}\ee

\subsection*{The Semi-Local Vortex Theory}

We would now like to discuss the low-energy dynamics of the vortex string. 
Vortices in gauge theories with $N_f>N_c$ are known as ``semi-local vortices'', 
terminology which first arose in the abelian gauge theory with multiple 
Higgs fields \cite{vash}. As the gauge coupling $e^2$ is varied, these 
solitons interpolate between Nielsen-Olesen like vortices and sigma-model 
lumps on the Higgs branch of the theory. In non-abelian theories of 
the type considered here, they were studied in \cite{vib}.

\para
Semi-local vortices involve a subtlety not shared by those  
discussed in Section 3: some of their zero modes are non-normalisable 
\cite{nonnorm}. This means that the Manton metric on their moduli space includes 
some (logarithmically) divergent terms and these fluctuations are classically 
frozen. The non-normalisability of these modes also leads to subtleties in 
treating these objects quantum mechanically which, to our knowledge, have 
not been resolved in the literature. 

\para
In \cite{vib}, a brane construction of both vortices and semi-local vortices 
was employed to extract the low-energy dynamics of the solitons. Although 
the resulting theory captured much information about vortex dynamics, 
it did not give the Manton metric on the moduli space. Indeed, for the case 
of semi-local vortices this discrepancy is most extreme since the brane construction 
provides 
a finite metric on the moduli space of semi-local vortices. 
Nevertheless, it was argued in \cite{vib} 
that as long as we restrict to BPS sectors of a supersymmetric gauge theory, 
then one should be able to use any of a class of metrics on the vortex moduli space since 
the questions reduce to calculating certain topological quantities. Here 
we present an example of this technique. Rather than using the non-normalisable metric on the 
semi-local vortex moduli space, we instead work with the simpler metric derived 
from the brane construction of \cite{vib}. The fact that we are able to reproduce 
the quantum spectrum of the four-dimensional gauge theory gives 
strong support in favour of this procedure. With this caveat in mind, we now 
describe the low-energy dynamics of the semi-local vortex \cite{vib}

\para
{\bf Semi-Local Vortex Theory:} {\it $d=1+1$, ${\cal N}=(2,2)$ supersymmetric $U(1)$ with a single
neutral chiral multiplet $Z$, $N_c$ chiral multiplets $\Psi_i$ of charge $+1$ 
and $(N_f-N_c)$ charged chiral multiplets $\tilde{\Psi}_m$ of charge $-1$. 
The $\Psi_i$ have twisted masses $m_i$, $i=1,\dots, N_c$, while the 
$\tilde{\Psi}_m$ have twisted masses $m_{N_c+m}$, $m=1,\ldots,N_f-N_c$.}

\para
The FI parameter is given by $r=2\pi/e^2$ as in Section 2 and 
the D-term for the theory reads,
\be
D=\sum_{i=1}^{N_c}|\psi_i|^2-\sum_{m=1}^{N_f-N_c}|\tilde{\psi}_m|^2-r
\nn\ee
After dividing by the $U(1)$ gauge action, the equation $D=0$ defines the 
Higgs branch of the theory for which, for $m_i=0$, is isomorphic to the 
internal moduli space of semi-local vortices \cite{vib}. Note that, in 
contrast to the ${\bf CP}^{N_c-1}$ of Section 2, the moduli space of semi-local vortices 
is non-compact. This reflects the fact that at large distances they look 
like sigma-model lumps, replete with a scaling modulus. When the masses 
are turned on $m_i\neq 0$, there are again only $N_c$ isolated vacua in 
the theory, given by $|\psi_j|^2=v^2\delta_{ij}$ and $\tilde{\psi}_m=0$. 
Once again, these correspond to the $N_c$ possible $U(1)$ embeddings of the Nielsen-Olesen 
abelian vortex. 

\para
The semi-local vortex theory described above was previously studied in 
\cite{dht} where it was shown that the exact BPS spectrum indeed 
coincides with the spectrum of massive quarks and monopoles in the 
four-dimensional parent theory. Once again, we will be brief and make only a 
few choice comments. As in Section 3, the anomalies in four and two 
dimensions are in agreement: for vanishing masses the $U(1)_R$ symmetry 
is broken by instantons to ${\bf Z}_{2(2N_c-N_f)}$ in both cases. 
The one-loop logarithmic running of the FI 
parameter is given by,
\be
r(\mu)=r_0-\frac{2N_c-N_f}{2\pi}\log\left(\frac{M_{UV}}{\mu}\right)
\label{runforestrun}\ee
which, agrees with the one-loop beta function of $e^2(\mu)$ in 
four-dimensions. Notice in particular that both two and four dimensional 
theories are asymptotically free for $N_f<2N_c$ and infra-red free 
for $N_f>2N_c$. Of particular interest is the critical case, 
$N_f=2N_c$. On the four-dimensional Coulomb branch, with vanishing masses, 
the theory is conformal. Once we move onto the Higgs branch, the 
same is true of the theory of the vortex string. It may prove 
interesting to understand the relevance of this point. 

\para
 Finally, the computation of the classical and quantum spectrum 
proceeds much as above -- for full details see \cite{dht} -- and reduces 
to computing the critical points of the effective superpotential,
\be
{\cal W}  = \frac{\tau\Sigma}{2} - \frac{1}{4\pi}\sum_{i=1}^{N_c}
(\Sigma-m_i)\log\left(\frac{2}{\mu}(\Sigma-m_i)\right) 
+\frac{1}{4\pi}\sum_{m=N_c+1}^{N_f}
\hspace{-0.3cm}(\Sigma-m_i)\log\left(\frac{2}{\mu}(\Sigma-m_i)\right) 
\nn\ee
The quantum corrected central charge takes the form \eqn{newz}, 
now with the $m_{D\,i}=-2i{\cal W}(e_i)$ where $e_i$ are the $N_c$ 
critical points of ${\cal W}$. Using the form of the superpotential 
above, we see that $m_{D\,i}$ is indeed given by \eqn{doogiehowsermd}. 
The exact BPS spectrum of the vortex string is in agreement with the 
BPS spectrum of its four-dimensional parent theory. Once again, 
the kinks have the interpretation of confined monopoles, while the 
elementary excitations of the string correspond to $N_c^2$ bound 
W-bosons, as well as $N_f(N_f-N_c)$ quark-string threshold states.

\section*{Appendix A: The Brane Construction}

In this section we review the brane derivation of the vortex theory 
given in \cite{vib} and present the (trivial) generalisation to include 
non-zero masses. While the construction of \cite{vib} was performed 
in the IIB string theory set-up of \cite{hw}, resulting in vortices 
as particles in $d=2+1$ dimensions, here we work with the T-dual IIA 
construction where the vortices appear as strings in a $d=3+1$ dimensional 
gauge theory. Related brane constructions of vortices were recently 
discussed in \cite{jeff}.

\para
Our brane configuration is drawn in Figure 2. We use the well-known construction 
of ${\cal N}=2$ theories in $d=3+1$ dimensions realised on the worldvolume of 
$N_c$ D4-branes suspended between two NS5-branes \cite{witm5}. A further $N_f=N_c$ 
D6-branes give rise to hypermultiplets coming from $4-6$ strings. The spatial worldvolume 
directions of the branes are
\be
NS5: && 12345 \nn\\
D4: && 1236 \nn\\ 
D6: && 123789 \nn\\
D2: && 39
\nn\ee
The gauge coupling $e^2$ and FI parameter $v^2$ are encoded in the separation 
$\Delta x$ of the two NS5-branes,
\be
\frac{1}{e^2}=\frac{\Delta x^6}{(2\pi)^2 g_sl_s}\ \ \ \ \ ,\ \ \ \ \ \ 
v^2=\frac{\Delta x^9}{(2\pi)^3g_sl_s^3}
\label{itsamiracle}\ee
\begin{figure}[htbp]
\begin{center}
\epsfxsize=6.0in\leavevmode\epsfbox{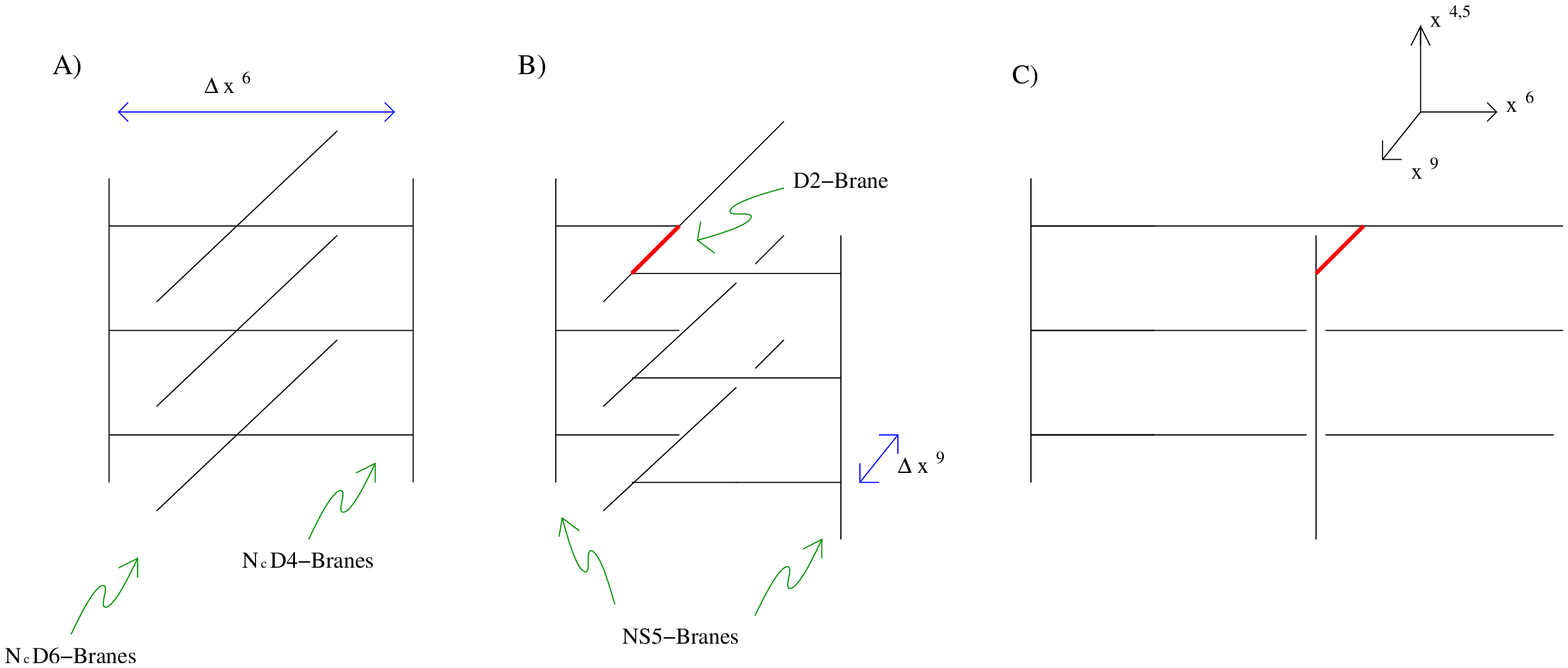}
\end{center}
 {\small Figure 2: The type IIA brane set-up. Figure A) shows the four dimensional theory at 
the root of the baryonic Higgs branch. In Figure B), the right-hand NS5-brane 
has slid in the $x^9$ direction and the gauge theory sits on the Higgs branch. We have 
included the D2-brane vortex string in red in this picture. 
In Figure C), we have moved the D6-branes 
off the page to the far-right, allowing us to read off the theory on the D2-brane.}
\end{figure}
where $g_s$ and $l_s=\sqrt{\alpha\prime}$ are the string coupling and string 
length respectively. The hypermultiplet masses and the vacuum expectation value 
of $\phi={\rm diag}(\phi_1,\ldots,\phi_{N_c})$ are 
encoded in the $x^4$ and $x^5$ positions of the D-branes \cite{witm5}
\be
m_i=\left.\frac{x^4+ix^5}{l_s^2}\right|_{D6_i}\ \ \  \ \ \ \ \ ,\ \ \ \ \  \ \ \ 
\phi_i=\left.\frac{x^4+ix^5}{l_s^2}\right|_{D4_i}
\label{d6mass}\ee
In Figure 2A) we draw the 
brane configuration corresponding to the four dimensional theory 
with $v^2=0$ at the root of the baryonic Higgs branch $\phi={\rm diag}(m_1,\ldots,m_{N_c})$. 
In Figure 2B), we have turned on the FI parameter $v^2$ by moving the right-hand 
NS5-brane out of the page in the $x^9$ direction. Here we also depict the vortex 
string, appearing as a D2-brane stretched the distance $\Delta x^9$ between the 
NS5-brane and the D3-brane. 

\para
To read off the vortex theory on the D2-brane, we first manipulate the branes a little. 
The field theory cares nothing for the $x^6$ position of the D6-branes and we may 
freely move them in this direction. Ther is one caveat however: they have non-zero linking 
number with the NS5-branes which ensures that D4-branes are created or destroyed if the 
two pass through each other \cite{hw}.
We choose to move the D6-branes to the right. 
When they pass through the right-hand NS5-brane, the connecting D4-branes dissapear 
by flux conservation and the D6-branes are now attached only to the left-hand NS5-brane. 
After moving the D6-branes to $x^6\rightarrow\infty$, the resulting configuration is 
shown in Figure 2C. From this we may read off the gauge theory on the D2-brane as 
described in \cite{hh}. It is given by $d=1+1$, ${\cal N}=(2,2)$ $U(1)$ gauge 
theory. The gauge coupling constant $g^2$ and the FI parameter $r$ are given by 
the separation of the NS5-branes,
\be
\frac{1}{g^2}=\frac{\Delta x^9l_s}{g_s}\ \ \ \ \  \ \ \ ,\ \ \ \  \ \ \ \ 
r=\frac{\Delta x^6}{2\pi g_sl_s}
\nn\ee
\begin{figure}[htbp]
\begin{center}
\epsfxsize=1.5in\leavevmode\epsfbox{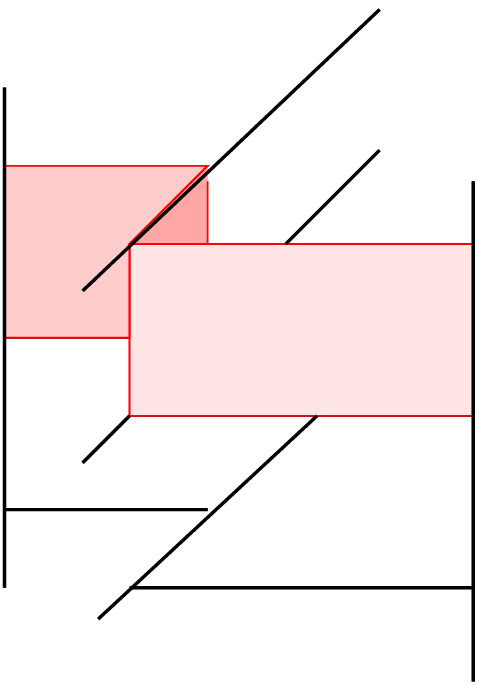}
\end{center}
 {\small Figure 3: The confined monopole in the brane picture. The D2-brane 
worldsheet interpolates between the upper and middle D4-branes as 
$-\infty < x^3 < +\infty$.}
\end{figure}
As explained in \cite{vib}, taking the decoupling limit of the four-dimensional 
gauge theory from the full string dynamics translates to the requirement that 
$g^2\rightarrow\infty$. In contrast, $r$ remains finite and, comparing with 
\eqn{itsamiracle}, is given by $r=2\pi/e^2$ as promised. The matter content of 
the D2-brane theory includes a single free chiral multiplet, corresponding to motion in the 
$x^1+ix^2$ direction, and $N_c$ charged chiral multiplets arising from the $2-4$ strings. 
These chiral multiplets have a twisted mass given by the position of the D4-branes 
\cite{hh},
\be
m_i=\left.\frac{x^4+ix^5}{l_s^2}\right|_{D4_i}
\nn\ee
which, for our choice of the baryonic Higgs branch, coincides with the hypermultiplet 
masses \eqn{d6mass}. This concludes the brane derivation of the vortex theory 
discussed in Section 2. 
\para
From the brane picture, certain other aspects of the vortex dynamics become  
immediately obvious. In Figure 2B), we drew the D2-brane attached to the upper 
D4-brane. This corresponds to a vortex string with magnetic flux in 
$B={\rm diag}(1,0,\ldots,0)$. It is clear from the brane picture that there 
exist a further $N_c-1$ inequivalent vortex configurations in which the 
D2-brane is attatched to one of the other D4-branes. It is also simple to understand 
the confined monopole in this picture. We consider a D2-brane worldvolume which 
starts attatched to the upper D4-brane at $x^3\rightarrow -\infty$, and then 
interpolates down to the middle D4-brane as $x^3\rightarrow +\infty$. At 
intermediate steps, the D2-brane cannot simply be a line stretching distance 
$\Delta x^9$ as drawn in Figure 2B) since it has no where to end. The only 
possiblity is that the D2-brane bends in the $x^6$ direction to attatch 
itself to the NS5-branes. The final configuration is drawn in Figure 3 and is similar 
to those considered in \cite{hh,dht}\footnote{Similar brane configurations have been 
considered by J. Evslin in the context of \cite{auzzi}.} Notice that as $v^2\rightarrow 0$, 
and the separation $\Delta x^9$ of the NS5-branes vanishes, this stretched D2-brane 
indeed becomes the 't Hooft Polyakov monopole in the Coulomb phase.

\section*{Appendix B: Potential on the Vortex Moduli Space}

One of the key features of the vortex theory described in Section 3 is 
the presence of twisted masses for the chiral multiplets. In Appendix 
A we saw how these arise from a brane construction and how they 
result in a potential on the moduli space of vortices which arises as the 
Higgs branch of the vortex theory. Here we provide a purely field theoretic 
derivation of this potential. The method follows closely that developed in 
\cite{me}.

\para
We are interested in non-abelian 
vortices in the four-dimensional theory described in Section 2. If 
the hypermultiplet masses vanish $m_i=0$, we may simply set the 
adjoint scalar $\phi=0$ and study the vortex equations \eqn{harry}. 
The question we wish to answer here is how these solutions are 
lifted with the introduction of the masses $m_i$. To simplify matters, 
we take all $m_i$ to be real, which allows us to restrict to real $\phi$ 
(the generalisation to complex masses is simple). Further, we will 
use the ability to shift $\phi$ to set $\sum_{i=1}^{N_f}m_i=0$. 
A solution to the vortex equations \eqn{harry} now has an extra contribution 
to its energy coming from the terms in the four dimensional action
\be
V=\int\ d^2x\ \frac{2}{e^2}{\rm Tr}\,{\cal D}_z\phi{\cal D}_{\bar{z}}\phi + 
\sum_{i=1}^{N_f}q_i^\dagger\,(\phi-m_i)^2 q_i
\nn\ee
which is to be evaluated on a particular configuration for the fields 
$A_z$ and $q_i$ solving \eqn{harry}. While $q_i$ and 
$A_z$ are fixed, $\phi$ may vary so as to minimize $V$. It satisfies,
\be
{\cal D}^2\phi=e^2\sum_{i=1}^{N_f}\ \{\phi,q_iq_i^\dagger\} - 2q_iq_i^\dagger m_i
\label{number1}\ee
subject to the asymptotic condition $\phi\rightarrow{\rm diag}(m_1,\ldots,m_{N_c})$. 
In this appendix we show how to evaluate $V$ for a given vortex solution. 

\para
The most general solution to the non-abelian vortex equations has $2kN_c$ 
parameters where $k$ is the magnetic flux \cite{vib}. Let ${\cal V}_{k,N_c}$ 
denote the moduli space of solutions and choose coordinates $X^p$ on 
\vkn\ 
with $p=1,\ldots, 2kN_c$. The tangent vectors of \vkn\ are provided by the 
zero modes $(\delta_p A_z, \delta_p q_i)$ 
of the vortex which satisfy the linearised version of \eqn{harry},
\be
{\cal D}_z\,\delta_p A_{\bar{z}}-{\cal D}_{\bar{z}}\,\delta_pA_z 
&=&\frac{ie^2}{2}\sum_{i=1}^{N_f}\left(\delta_pq_i\,q_i^\dagger+q_i\,\delta_pq_i^\dagger\right)
\nn\\
{\cal D}_z\,\delta_pq_i&=&i\delta_pA_z\,q_i
\label{mamta}\ee
This is to be augmented by the gauge fixing condition arising from Gauss' law
\be
{\cal D}_z\,\delta_pA_{\bar{z}}+{\cal D}_{\bar{z}}\,\delta_pA_z=-\frac{ie^2}{2}
\sum_{i=1}^{N_f}\left(\delta_pq_i\,q_i^\dagger-q_i\,\delta_pq_i^\dagger\right)
\label{fixit}\ee
The Manton metric on \vkn\ is defined by the overlap of zero modes,
\be
g_{pq}= \int d^2x\ \frac{2}{e^2}{\rm Tr}(\delta_{\{p}A_z)\,(\delta_{q\}}A_{\bar{z}})
+\sum_{i=1}^{N_f}(\delta_{\{p}q_i)\,(\delta_{q\}}q_i^{\dagger})
\label{manton}\ee
Of particular interest will be the zero modes generated by symmetries, specifically 
the action of the $SU(N)_{\rm diag}$ symmetry preserved by the vacuum when 
$m_i=0$. As we have seen in Section 3, for the case of a single vortex $k=1$, 
this sweeps out the entire ${\bf CP}^{N_c-1}$ internal vortex moduli space 
\cite{vib,yung}. For higher $k$, it provides only a subset of the zero modes. 
In all cases, the action of the symmetry results in an $SU(N_c)_{\rm diag}$ 
isometry of the moduli space metric $g_{pq}$ 
with $N_c-1$ mutually commuting holomorphic 
Killing vectors. These will be important in the following.
As explained in Section 2 of \cite{vib}, the zero modes associated with this 
symmetry can be constructed 
uniquely from a given  Lie algebra element $\Omega_0\in su(N_c)_{\rm diag}$. 
The zero modes are given by,
\be
\delta A_z = {\cal D}_z\Omega\ \ \ \ \ ,\ \ \ \ \ \ \ \delta q=i(\Omega q -q\Omega_0)
\label{zeromodes}\ee
where $\Omega=\Omega(z,\bar{z})$, a function which, from \eqn{mamta} and \eqn{fixit}, 
satisfies,
\be
{\cal D}^2\Omega = e^2\sum_{i=1}^{N_f}\ \{\Omega,q_iq_i^\dagger\} - 2q_iq_i^\dagger \Omega_0
\label{number2}\ee
subject to the boundary condition $\Omega(z,\bar{z})\rightarrow \Omega_0$ as 
$|z|\rightarrow\infty$. Now let us choose a very special element $\Omega_0$ 
which lies in the Cartan subalgebra of $su(N_c)_{\rm diag}$. We set
\be
\Omega_0={\rm diag}(m_1,\ldots,m_{N_c})
\label{omego}\ee
The crucial observation is that for this specific rotation, equation \eqn{number2} 
coincides with the equation of motion for $\phi$ given in \eqn{number1}: we have  
$\Omega=\phi$.  This allows us to write the excess energy of the vortices in 
terms of the overlap of these zero modes \eqn{zeromodes}
\be
V=\int d^2x\ \frac{2}{e^2}{\rm Tr}\ \delta A_z\,\delta A_{\bar{z}} 
+\sum_{i=1}^{N_f} \delta q_i\delta q_i^\dagger
\label{vend}\ee
We are now almost done. The final step is to decompose the specific rotation 
\eqn{omego} into a basis of normalised rotations. It is somewhat simpler 
to work with the larger $u(N_c)_{\rm diag}$ Cartan sub-algebra and subsequently 
impose the vanishing trace condition on various objects. Let 
${H_i}$ denote the $N_c$ mutually commuting generators and write 
$\Omega_0 = \sum_i m_i H_i$. We further denote by $K_i$ the Killing 
vector on \vkn\ which is generated by the action of $H_i$. Note that, 
because of the traceless condition,  these $N_c$ Killing vectors are not 
all linearly independent but satisfy $\sum_iK_i=0$. We may now express 
the zero mode \eqn{zeromodes} in this basis of tangent vectors,
\be
\delta A_z = \left(\sum_{i=1}^{N_c} m_iK_i^p\right)\,\delta_pA_z
\ \ \ \ \ ,\ \ \ \ \ \ \ \delta q=\left(\sum_{i=1}^{N_c}m_iK_i^p\right)\,\delta_pq
\nn\ee
Finally, inserting this into \eqn{vend} and using the defintion of the metric 
\eqn{manton}, we arrive at our promised result for excess vortex energy as a 
potential on \vkn\ given by the 
length-squared of a particular Killing vector,
\be
V= \sum_{i,j=1}^{N_c} (m_iK_i^p)\,(m_jK_j^q)\,g_{pq}
\label{yippee}\ee

\subsection*{From Vortex Theory to Vortex Moduli Space}

The vortex theory described in Section 2 (and derived in Appendix A 
using branes) is given in terms of a gauged linear sigma-model. The Higgs 
branch of the vortex theory coincides with the moduli space of vortices which, 
for a single vortex $k=1$, is simply ${\bf C}\times {\bf CP}^{N_c-1}$. 
Here we would like to show how the potential \eqn{yippee} arises 
from the twisted mass terms in the vortex theory. 
The Higgs branch is defined by the D-term constraint 
\be
D= \sum_{i=1}^{N_c}|\psi_i|^2-r = 0
\nn\ee
modulo the $U(1)$ gauge action which rotates each chiral multiplet equally: 
$\delta_{\rm gauge}\psi_i=i\psi_i$. The Higgs branch inherits a natural 
metric from the gauge theory through a mechanism known as the K\"ahler 
quotient. In the present context, this is simply the round Fubini-Study metric 
on ${\bf CP}^{N_c-1}$. The metric on the Higgs branch is defined  
in terms of a basis of tangent vectors $\delta_p\psi_i$, $p=1,\ldots, 2(N_c-1)$ 
satisfying the linearised equations $\delta_pD=0$ together with the gauge 
fixing constraint $\sum_{i} \psi_i^\dagger\,\delta_p\psi_i=0$. The metric 
on the Higgs branch is then given by
\be
g_{pq}=\sum_{i=1}^{N_c}(\delta_{\{p}\psi_i)\,(\delta_{q\}}\psi^\dagger_i)
\label{higgsmet}\ee
For the single $k=1$ vortex considered here, all directions of 
the inernal moduli space are generated by the action of the $SU(N_c)_{\rm diag}$ 
symmetry\footnote{For the more general $k>1$ theories discussed 
in \cite{vib}, this statement is no longer true but the following 
methods can also be implemented.}. We will match to the vortex 
moduli space calculation described above by following the action of this 
symmetry and, in particular, the $(N_c-1)$ mututally commuting Killing vectors 
it generates on the Higgs branch.
As above, it will prove useful to overcount and work with $N_c$ Killing 
vectors subject to a constraint. 
Consider the $N_c$ normalised zero modes arising from such the $su(N_c)_{\rm diag}$ 
action.
\be
\delta_i\psi_j=i\psi_j\delta_{ij} - \frac{i}{r} |\psi_i|^2\psi_j
\label{killbill}\ee
These are not all linearly independent. If we denote the corresponding Killing 
vector on the Higgs branch as $K_i$, we have $\sum_{i=1}^{N_c}K_i=0$. Note that 
the action has been normalised so that $K_i$ coincides with the Killing vector on 
${\cal V}_{1,N_c}$ defined above. It is simple to see how the masses $m_i$ 
affect the Higgs branch. In the strict $g^2\rightarrow \infty$ limit, they 
induce a potential given by the term from equation \eqn{ours}
\be
V=\sum_{i=1}^{N_f}|\psi_i|^2(\sigma-m_i)^2
\label{vyes}\ee
where $\sigma$ can vary so as to minimise $V$, giving rise to 
the solution
\be
\sigma=\frac{1}{r}\sum_im_i|\psi_i|^2
\nn\ee
Substituting this into the potential \eqn{vyes}, we see that we can express 
$V$ purely in terms of geometrical objects on the Higgs 
branch: the metric \eqn{higgsmet} and the Killing vectors $K_i$ arising 
from the action \eqn{killbill}. We have,
\be
V&=&\sum_{i=1}^{N_c}|m_i|^2|\psi_i|^2-\frac{1}{r}|\sum_im_i|\psi_i|^2|^2 
\nn\\ &&
=\sum_{i,j=1}^{N_c} (m_iK_i^p)\,(\bar{m}_jK_j^q)\, g_{pq}
\nn\ee
in agreement with the expression \eqn{yippee}. It is heartwarming that 
the potentials derived from field theory and branes coincide.

\newpage

\subsection*{The Acknowledgments}
This work was first presented by D.T. as part of the COE lectures on 
``Instantons, Vortices, Monopoles and Kinks'' at Tokyo Institute of Technology, 
February 2004. D.T. would like to thank Norisuke Sakai for his kind invitation and 
all members of TITech for their valuable contribution during the lectures. We are 
also grateful to Misha Shifman and Alyosha Yung for providing us with a preliminary version 
of \cite{misha} and for pointing out a mistake regarding the BPS quark masses in the 
original version of this paper. 
A.H. is supported in part by the Reed Fund Award and a DOE OJI Award.
D.T. is supported by a Pappalardo fellowship and is grateful to   
the Pappalardo family for their generosity. This work was also supported in part 
by funds provided by the U.S. Department of Energy (D.O.E.) under 
cooperative research agreement \#DF-FC02-94ER40818.

\end{document}